\documentclass[twocolumn,showpacs,preprintnumbers,superscriptaddress,amsmath,amssymb,floatfix,prl]{revtex4}

\usepackage{graphicx}
\usepackage{dcolumn}
\usepackage{bm}
\usepackage[utf8]{inputenc}
\usepackage{color}

\begin{document}

\preprint{}

\title{Joule-level high effiency energy transfer to sub-picosecond laser pulses\\ by a plasma-based amplifier}

\author{J.-R. Marqu\`es}
\email[corresponding author: ]{jean-raphael.marques@polytechnique.fr}
\affiliation{LULI, CNRS, \'Ecole Polytechnique, CEA, Universit\'e Paris-Saclay, Sorbonne Universit\'e, F-91128 Palaiseau, France}
\author{L. Lancia}
\email[{\it SBAI, Universit\`a di Roma, La Sapienza, via Scarpa 14, 00161 Rome, Italy} at the time of the experiment.]{}
\affiliation{LULI, CNRS, \'Ecole Polytechnique, CEA, Universit\'e Paris-Saclay, Sorbonne Universit\'e, F-91128 Palaiseau, France}
\author{T. Gangolf}
\affiliation{LULI, CNRS, \'Ecole Polytechnique, CEA, Universit\'e Paris-Saclay, Sorbonne Universit\'e, F-91128 Palaiseau, France}
\affiliation{ILPP, Heinrich-Heine Universit{\"a}t D{\"u}sseldorf, 40225 D{\"u}sseldorf, Germany}
\author{M. Blecher}
\affiliation{ILPP, Heinrich-Heine Universit{\"a}t D{\"u}sseldorf, 40225 D{\"u}sseldorf, Germany}
\author{S. Bola{\~n}os}
\affiliation{LULI, CNRS, \'Ecole Polytechnique, CEA, Universit\'e Paris-Saclay, Sorbonne Universit\'e, F-91128 Palaiseau, France}
\author{J. Fuchs}
\affiliation{LULI, CNRS, \'Ecole Polytechnique, CEA, Universit\'e Paris-Saclay, Sorbonne Universit\'e, F-91128 Palaiseau, France}
\author{O. Willi}
\affiliation{ILPP, Heinrich-Heine Universit{\"a}t D{\"u}sseldorf, 40225 D{\"u}sseldorf, Germany}
\author{F. Amiranoff}
\affiliation{LULI, Sorbonne Universit\'e, CNRS, \'Ecole Polytechnique, CEA, Universit\'e Paris-Saclay, F-75252 Paris, France}
\author{R. L. Berger}
\affiliation{Lawrence Livermore National Laboratory, Livermore, California 94550, USA}
\author{M. Chiaramello}
\affiliation{LULI, Sorbonne Universit\'e, CNRS, \'Ecole Polytechnique, CEA, Universit\'e Paris-Saclay, F-75252 Paris, France}
\author{S. Weber}
\affiliation{Institute of Physics of the ASCR, ELI-Beamlines, 18221 Prague, Czech Republic}
\author{C. Riconda}
\affiliation{LULI, Sorbonne Universit\'e, CNRS, \'Ecole Polytechnique, CEA, Universit\'e Paris-Saclay, F-75252 Paris, France}

\date{\today}

\begin{abstract}
Laser plasma amplification of sub-picosecond pulses above the Joule level is demonstrated, a major milestone for this scheme to become a solution for the next-generation of ultra-high intensity lasers. By exploring over 6 orders of magnitude the influence of the incident seed intensity on Brillouin laser amplification, we reveal the importance of a minimum intensity to ensure an early onset of the self-similar regime, and a large energy transfer with a very high efficiency, up to 20\%. Evidence of energy losses of the seed by spontaneous backward Raman is found at high amplification. The first three-dimensional envelope simulations of the sub-picosecond amplification were performed, supplemented by one-dimensional PIC simulations. Comparisons with the experimental results demonstrate the capability of quantitative predictions on the transferred energy. The global behavior of the amplification process, is reproduced, including the evolution of the spatial profile of the amplified seed.
\end{abstract}


\maketitle

\section{I. INTRODUCTION}
Laser amplification in a plasma is emerging as a real alternative to solid state optics when it comes to manipulating light at very high intensities, due to the possibility to overcome the limits imposed by the damage threshold of materials along with the promising perspectives of compactness and low cost. The main concerns in laser plasma amplification are about the experimental feasibility, the extent of parameter ranges of applicability, the competing sources of energy dissipation, and the efficiency of the process. The way the plasma responds to an external excitation and couples to it, fully characterizes the performances of a plasma amplifier. The schemes proposed and studied so far exploiting an electron or ion response of the plasma (resonant Raman \cite{Vieux2017, Ren, Wu2018},  chirped pump Raman \cite{Yang2015, Ersfeld2005}, Raman with flying focus \cite{Turnbull2018}, diverging Raman \cite{Sadler2018}, multi-frequency Raman \cite{Barth2018}, plasma wave seed for Raman \cite{Qu2017}, single-pulse Raman \cite{Malkin}, strongly-coupled Brillouin \cite{Edwards2016}, single-pulse Brillouin \cite{Peng2016}, few-cycle laser amplification by Brillouin \cite{Zhang}) have explored different parameter ranges and beam configurations in order to optimize (or minimize) a particular aspect of the amplification. Each scheme, according to which type of plasma excitation is exploited, has been put forward as the most appropriate for a specific purpose. In particular, the energy coupling between two beams (transverse electromagnetic waves) at the same frequency through forced ion plasma oscillations (Brillouin amplification in the strong coupling regime) has been considered due to the simpler experimental setup (2 beams produced by the same oscillator) and the robustness it offers in terms of frequency mismatch \cite{Amiranoff}.
The energy amplification of a sub-picosecond seed beam, mediated by a forced ion response of the plasma, has been proposed \cite{Andreev2006} and experimentally demonstrated \cite{Lancia2010}. The optimal parameter range has been assessed \cite{Chiaramello2016a, Edwards2016}, together with the role of the chirp of the laser \cite{Lehmann2015, Chiaramello2016b, Amiranoff}. The various phases of the amplification process have been quantified \cite{Schluck, Amiranoff} and the transition to the efficient self-similar regime has been  experimentally demonstrated \cite{Lancia2016}. The interest in this topic extends also to energy transfer in the nanosecond regime in the context of inertial confinement fusion \cite{Kirkwood2018a, Kirkwood2018b}. Among the schemes cited above, many of them, theoretically and/or numerically explored, have predicted very high efficiency. For example, amplification of a 0.3mJ seed pulse above 1.2 J is predicted using the diverging scheme \cite{Sadler2018}, while a 15 J - 50 fs could be amplified to 5 kJ - 60 fs from a 10  kJ -1 ns pump in a optimized plasma amplifier \cite{Sadler2017}. However, experimental results are very rare and the transferred energies and efficiencies are much lower. To our knowledge, the maximum energy transfer yet reported is 170 mJ from a 70 J pump laser via the Raman scheme \cite{Vieux2017}, corresponding to an efficiency of only 0.24 \%. From Brillouin amplification \cite{Lancia2010} 60 mJ has been transferred from a 2 J pump laser, corresponding to a 3\% total efficiency.
In this paper we demonstrate laser plasma amplification of sub-picosecond pulses above the Joule level, reaching an unprecedented energy transfer of up to 2 J, with a very high efficiency, up to 20\%. We present the first investigation of Brillouin laser amplification as a function of the incident seed intensity. Exploring 6 orders of magnitude, up to intensities above the pump one, we reveal the importance of using a high-enough incident intensity to enter the self-similar regime as promptly as possible and get a highly efficient energy transfer in a sub-picosecond scale. In this regime of efficient amplification, we observed evidence of energy losses of the seed by spontaneous backward Raman. Due to the crossed beam geometry of our setup, we also observe a modification of the spatial profile of the seed by the amplification process. The first three-dimensional envelope simulations of sub-picosecond laser-plasma amplification are presented, supplemented by one-dimensional PIC simulations. Quantitative agreement is obtained on the transferred energy. The global behavior of the amplification process is well reproduced, including the evolution of the spatial profile of the seed. We show that the main sources of loss of pump energy available for transfer, in particular spontaneous Raman and Brillouin Backscattering, can be controlled by an appropriate choice of the pump and seed timing as well as initial seed intensity.

\section{II. EXPERIMENTAL SETUP AND METHOD}
\begin{figure*}
\includegraphics[width=\textwidth]{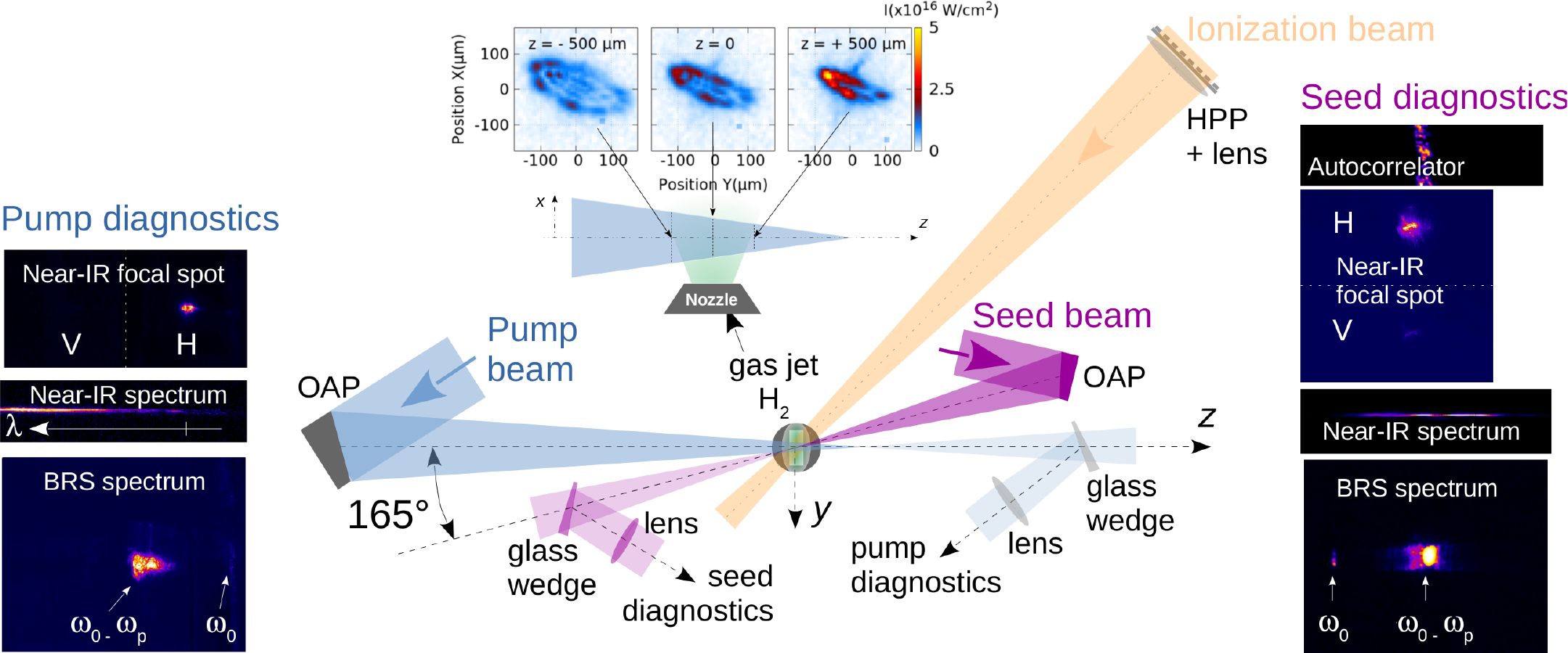}
\caption{\label{setup}Experimental arrangement, showing pump and seed beams focused by off-axis parabola (OAP), the ionization beam focused using a lens after a hybrid phase plate (HPP), the supersonic hydrogen gas jet at the beams intersection, the glass wedges sending a low intensity fraction of the beams to lenses imaging the beams focus toward their diagnostics, raw images of each diagnostics, and transverse profiles of the pump at 3 positions in the gas jet (z = -500, 0, and +500 $\mu$m from the plasma center). Pump and seed diagnostics are described in section II.}
\end{figure*}

The results presented here have been obtained at LULI on the ELFIE facility. A schematic view of the experimental setup is shown in Fig.~\ref{setup}. 
A “pump” and a “seed” beam, linearly polarized and at the same wavelength (1058 nm), crossed in a plasma at an angle of 165$^{\circ}$ in the $yz$-plane. The seed  was focused at its minimum focal spot $\sigma_x \times \sigma_y \simeq 100\times 60\, \mu{\rm m}^2$ (Full-Width at Half-Maximum, FWHM) and compressed to its minimum duration (FWHM $\sim$ 0.55 ps). 
Its maximum energy was $\sim$ 4 J, providing a maximum intensity of $\sim 10^{17}$ W/cm$^2$.
To increase the interaction volume, the 9 J pump was defocused and stretched, to a focal spot of $\sigma_x \times \sigma_y \simeq 85\times 135\, \mu{\rm m}^2$ (FWHM) and a duration of 1.7 ps FWHM (red before blue, $\omega(t) = \omega_0[1 - 2 \alpha \omega_0 t]$, $\alpha = -7.65\times 10^{-7}$), leading to a maximum intensity of $\sim 4 \times 10^{16}$ W/cm$^2$.
The plasma was created 1 ns before pump and seed arrival by a third beam (35 J - 500 ps - 1058 nm) propagating at 135$^\circ$ from the pump. This beam was focused to a $\sigma_x \times \sigma_z \simeq 200\times 1200\, \mu{\rm m}^2$ spot, reaching an average intensity of $\sim 2\times 10^{13}$ W/cm$^2$ that fully ionized a hydrogen supersonic gas jet. The resulting plasma density profile was Gaussian ($\sigma_y \times \sigma_z \simeq 750\times 575\, \mu{\rm m}^2$ FWHM) with a maximum electron density adjusted to  $n_e \simeq 4-5\times 10^{19}\, {\rm cm}^{-3} (\sim 0.04-0.05\, n_c$, where $n_c$ is the critical density at the laser wavelength). The electron and ion temperatures at pump and seed arrival were $T_e \sim$ 100 eV and $T_i \simeq$ 80 eV (from 2D fluid simulations with the code FCI2 \cite{Loiseau}). Let us note that in strong coupling regime, even if $T_e \approx T_i$, due to their large phase velocity Landau damping on ion perturbations will be relatively small. At the plasma density and temperature of the present experiment, the pump intensity threshold for the strongly-coupled regime is $I_{th} \sim 10^{13}$ W/cm$^2$, reached very early in the pump leading edge. The pump and seed waves are at the same frequency with a relatively narrow bandwidth ($\Delta\omega/\omega_0 \sim 5\times 10^{-3}$). The time delay $t_p-t_s$, where $t_p$ and $t_s$ are the arrival times of the maximum of the pump and seed beams at the plasma center, was adjusted with a delay line. For $t_p-t_s>0$, the seed arrives before the pump.

Let us note that a fully counter-propagating (180$^{\circ}$) geometry would have been better for an optimum coupling between pump and seed, as demonstrated in \cite{Lancia2016}. However, the incident seed intensity in \cite{Lancia2016} was $\sim 10^{13}$ W/cm$^2$ (few mJ). In the present experiment, to explore interaction at much higher incident seed intensities (up to $10^{17}$ W/cm$^2$, few Joules) without risking to damage the laser chain with counter-propagating beams, we had to use a 165$^{\circ}$ geometry, even if it reduces the interaction length, the quality of the transverse overlap, and the control of pump spontaneous Raman scattering.

The focal spots and spectra of the transmitted pump and seed beams near their central wavelength (1050-1070 nm) were recorded on high-dynamic CCD cameras. The infrared backward (165$^{\circ}$) Raman spectra (1100-1700 nm) of each beam were recorded on InGaAs CCD cameras, providing the amount of Raman signal and allowing the retrieval of the electron plasma density. In order to measure the transmitted pump and seed energies, the focal spot diagnostics were absolutely calibrated. The pulse duration of the transmitted seed was measured with a second order autocorrelator coupled to a high-dynamic CCD camera.\\
The growth rate of strongly-coupled Brillouin (responsible for laser amplification), as well as those of spontaneous Raman or filamentation (responsible for beam losses or deterioration) depend on plasma density. All the following results are at $n_e \sim 0.05 n_c$, where we measured the maximum energy transfer. Also, they all correspond to pump and seed pulses synchronized at the center of the plasma.\\

\section{III. EXPERIMENTAL RESULTS}
\subsection{A. Amplification as a function of seed incident energy.}
\begin{figure*}
\includegraphics[width=\textwidth]{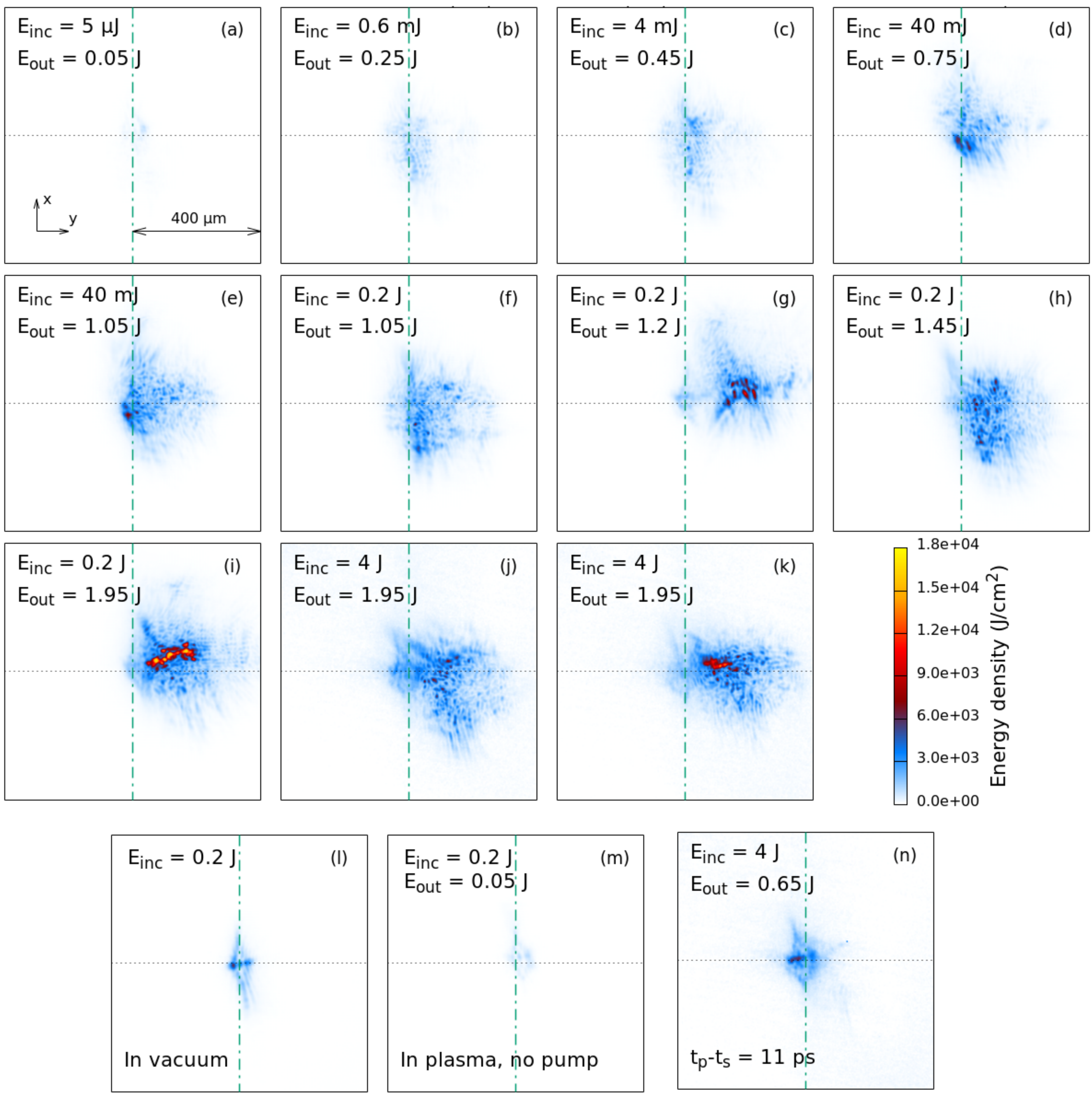}
\caption{\label{Foc_seed}Transverse profile of the transmitted seed pulse (energy per unit area of the focal spot), (a) to (k): for different incident seed energies, (l): in vacuum, (m): in plasma without pump, (n): in plasma with the pump arriving 11 ps after the seed ($t_p-t_s$ = +11 ps). Except for (n), pump and seed reach the center of the plasma at the same time ($t_p-t_s$ = 0). The spatial and color scales are the same for all the images, as well as the pump and plasma parameters.}
\end{figure*}

\begin{figure}
\resizebox{85mm}{!}{\includegraphics{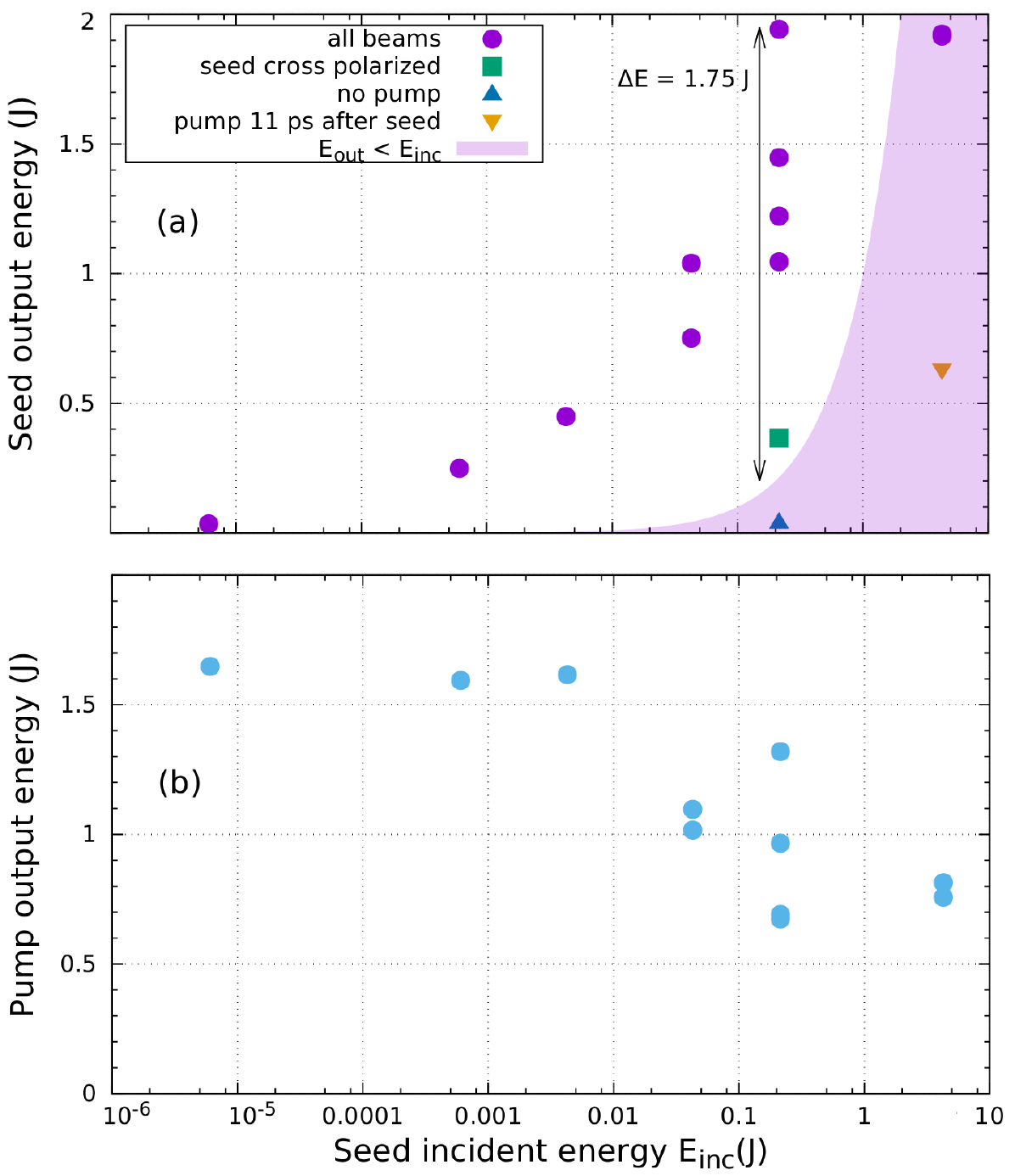}}
\caption{\label{Eout_vs_Eseed}Energy of the transmitted beams as a function of the incident seed energy $E_{inc}$, measured by integration of the energy calibrated focal spot images of the seed (a) and the pump (b). The corresponding incident seed intensity in vacuum is $I_{inc}$ (W/cm$^2$) $\sim 3.2\times 10^{16} E_{inc}$(J). Points out of the purple area in (a) correspond to seed amplification. At $E_{inc}$ = 0.2 J the energy gain is $\Delta E$ = 1.75 J. The green square and the orange triangle in (a) are two cases where there is no EM-coupling but other mechanisms are present (spontaneous scattering, hydrodynamics, etc ...).}
\end{figure}

The influence of the incident seed intensity on the final energy transfer (energy gain) has been studied by changing only the input seed energy ($5\times 10^{-6}$ to 4 J, half of the pump energy), all the other pump, seed and plasma parameters being fixed. Fig.\ref{Foc_seed} presents the seed focal spot at the output of the plasma, after interaction with the pump, for different incident seed energies.
For every image in Fig.\ref{Foc_seed}, the labels indicate the seed incident and output energies. These values are reported in Fig.\ref{Eout_vs_Eseed}, together with “test” shots: without the pump, or with orthogonal pump and seed polarizations. The filled area in Fig.\ref{Eout_vs_Eseed}a represents the separation between gain (outside the area) and loss (inside). From these two figures, one can see that when the incident seed energy is increased from $5\times 10^{-6}$ to 0.2 J ($I_{inc}$ from $1.6\times 10^{11}$ to $6.4\times 10^{15}$ W/cm$^2$), the amount of energy transferred from the pump to the seed increases up to a value of $\Delta E \sim$ 1.75 J, which is 20-25 \% of the total pump energy. Such an efficiency indicates that the interaction occurs in the pump depletion regime, as expected from self-similar Brillouin amplification. To our knowledge, this is an unprecedented level of energy transfer and efficiency from a laser-plasma amplifier of sub-picosecond pulses.

As expected from the three-wave coupling, when pump and seed propagate in plasma with orthogonal polarizations (green square), they cannot couple and the transferred energy drops. A small transfer is still observed due to the non-perfectly defined polarization of the beams.

When the seed incident intensity is larger than the pump one ($\geq$ few $10^{16}$ W/cm$^2$, corresponding to an incident seed energy $\geq$ 1 J), the efficiency of the amplification process drops. The seed transmission falls to 50 \%. It is interesting to compare this with a case where no coupling occurs (orange triangle, $t_p-t_s$ = 11 ps, pump late after seed): here transmission drops at 80\%. This indicates that even in this particular case of unfavorable initial coupling conditions, energy is still transferred from pump to seed.

In the self-similar regime the seed duration and intensity are correlated, the larger the intensity the smaller the duration \cite{{Chiaramello2016a},{Trines}}. Since the input seed duration in our experiment was fixed (550 fs), depending on its incident intensity the interaction did not start immediately in the self-similar regime which reduced the efficiency and slowed down the compression process. This non-optimal seed pulse was particularly deleterious near $10^{17}$ W/cm$^2$ where its initial duration was much too long for an optimum energy transfer and compression.

The interest of the crossed polarization and the $t_p-t_s$ = 11 ps cases compared to shots without pump is that in these configurations the EM-coupling is ineffective while the other phenomena affecting the pulses propagation are preserved (hydrodynamics, heating, spontaneous instabilities, etc ...), allowing to clearly separate them.
 
The fluctuations observed in the transferred energy are due to the shot to shot variations of laser-plasma parameters, mainly the seed and pump beam pointing stability. This was enhanced by the fact that, due to the experimental arrangement, when the pump moved toward one direction, the seed moved toward the opposite, strongly affecting the overlap.

The transmitted pump energy as a function of the incident seed energy is presented in Fig.\ref{Eout_vs_Eseed}b. One can see that the higher the seed energy gain (Fig.\ref{Eout_vs_Eseed}a), the lower the transmitted pump energy. At very weak incident seed (no energy transfer) the pump transmitted energy is still quite low, of the order of 1.7 J, which is about 20 \% of the incident energy. This relatively low transmission is due to beam refraction and to the losses by collisional absorption and spontaneous Raman instability, both favoured by the low plasma temperature (100 eV). On the contrary, when the seed is injected with a high-enough intensity, Brillouin is stimulated and competes with spontaneous Raman instability \cite{Lancia2016}.
Let us note that because of the losses undergone by the pump before crossing the seed, the available pump energy is between 9 J (vacuum) and 1.7 J (plasma exit), indicating that the amount of effective energy transfer at the interaction point is even higher than 25 \%.

The other diagnostics (pump and seed spectra, seed autocorrelator) were not calibrated in energy, but gave the same relative increase of the transferred energy with the injected seed energy, as well as the same relative pump depletion.

\begin{figure}
\resizebox{85mm}{!}{\includegraphics{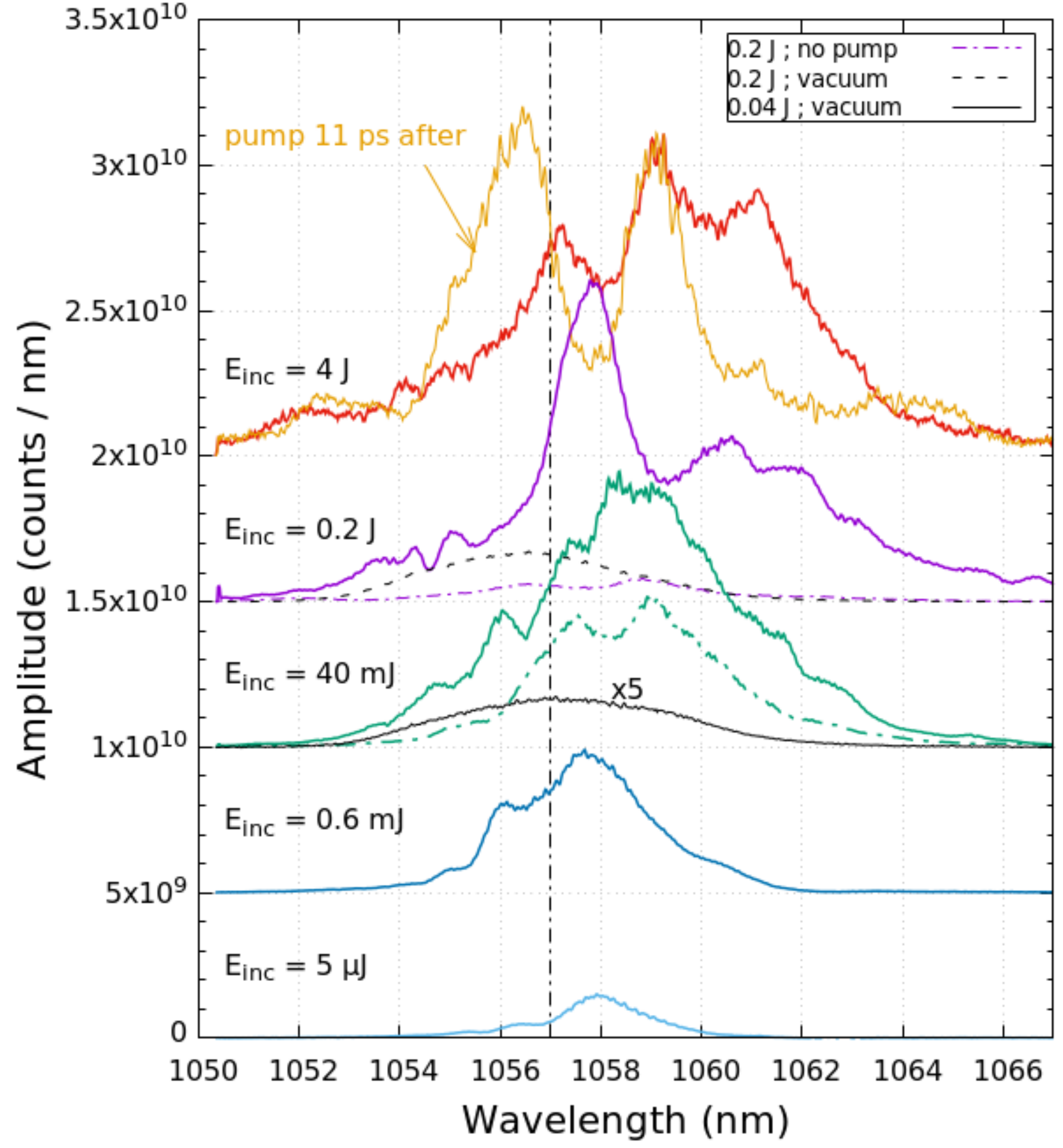}}
\caption{\label{Spectrum_vs_Eseed}Transmitted seed spectrum for different incident seed energies. Black lines are reference shots in vacuum. The spectrum for $E_{inc}$ = 40 mJ J in vacuum (black plain line) has been multiplied by a factor 5 for a better visibility. For the same purpose an offset of $5 \times 10^9$ is added to each set of curves.}
\end{figure}

The seed spectra are represented in Fig.\ref{Spectrum_vs_Eseed} at different energies of the injected seed. It shows that the amplification not only happens over the entire seed spectrum, but also that the higher the energy transfer, the broader and more red shifted is the amplified spectrum, a signature of SBS occurring in the self-similar regime \cite{Lancia2016}.

The two uppermost spectra in Fig.\ref{Spectrum_vs_Eseed} correspond to the case where the seed incident intensity is larger than the pump one and the energy transfer is not efficient. However, one can see that when the seed does not interact with the pump (yellow curve, pump 11 ps after seed), its spectrum remains centered at its initial vacuum position (1058 nm) and is not broadened. Moreover, a depletion is observed at the center of the spectrum. This is where the intensity is highest and the SRS and SBS spontaneous instabilities grow fastest. In opposite, when the pump overlaps and couples to the seed, it transfers some energy that compensates part of its losses (orange triangle versus violet circle in Fig. \ref{Eout_vs_Eseed}(a) at $E_{inc}$ = 4 J), refilling the seed spectrum (red curve in Fig.\ref{Spectrum_vs_Eseed}). Even if the losses are still larger than the energy gain, the interaction occurs in the self-similar regime, red-shifting and broadening its spectrum.

\subsection{B. Transverse profile of the amplified seed.}
The bottom images in Fig.\ref{Foc_seed} show the transmitted seed focal spot in vacuum (l), at the plasma exit without (m), and with (i) pump, at a same incident seed energy (0.2 J). Without the pump the beam undergoes a factor $\sim$ 5 attenuation but its transverse profile is not much different from the vacuum case. Conversely, with the pump and at high energy transfer ($\Delta E \sim$ 1.7 J, $\sim$ 20 \% of the pump energy), the amplification process strongly modifies the focal spot shape and increases its size. The evolution of the seed focal spot is due to: (i) the evolution of the pump profile along the interaction region, as shown in Fig.\ref{setup} at three positions (-500, 0, +500 $\mu$m) along the propagation axis; (ii) the 165$^{\circ}$ coupling geometry leading to a progressive transversal drift of the seed pulse through the pump focal spot; (iii) the amplification process that is non-linear with the incident seed intensity (Fig.\ref{Eout_vs_Eseed}). The resulting increase of the focal spot size thus reflects the very high efficiency of the amplification process, that makes maximum use of the available pump energy.

On the amplified focal spot one can observe very small and randomly distributed structures. At our typical pump intensity of $\sim 10^{16}$ W/cm$^2$, the filamentation growth rate in a hydrogen plasma at 0.05 $n_c$ is $\sim$ 2 ps, about 20 times slower than the sc-SBS growth rate. Filamentation will become significant after few growth rates, so few times more than the duration of our pump pulse (1.7 ps), or the time for the pulses to cross the plasma length ($\sim$ 2 ps). Since the plasma profile is Gaussian, it even takes more time on the lower density sides of the profile. This is confirmed by 2D-PIC simulations \cite{{Weber},{Riconda}} and by 3D-fluid simulations (see below). Moreover, our experimental results confirm these predictions. Indeed, the transmitted focal spot of the pump did not present the small-size, randomly-distributed and shot to shot changing pattern characteristic of a beam undergoing filamentation. This was also true for the seed beam when it was propagating in the plasma without the pump (Fig.\ref{Foc_seed}(m)(n)). In addition, shots with a pump duration or a plasma length doubled did not show any filamentation of the beam. The very small structures observed on the transmitted seed focal spot appear only when it is coupled to the pump and amplified. We attribute them to the initial spatio-temporal non uniformities in amplitude and phase of the pump, seed and plasma that couple during their propagation and generate this speckle-like pattern.

\subsection{C. Duration of the amplified pulse.}
\begin{figure*}
\includegraphics[width=\textwidth]{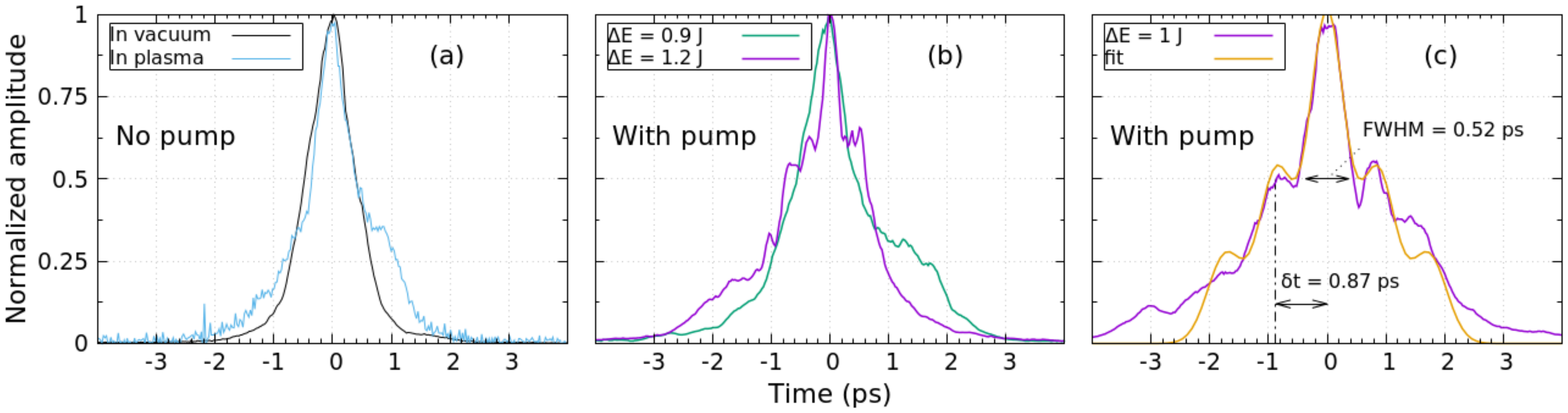}
\caption{\label{Autoco}Autocorrelation traces of the transmitted seed pulse, normalized to their maximum to help the comparison of their width. (a): without the pump, under vacuum and in plasma ; (b): in plasma with pump, for different transferred energies. (c): in plasma with pump, case with “$\pi$-pulses”. The orange curve is a fit using a triple Gaussian function: $e^{-(t/\tau)^2} + A [e^{-(t+\delta t)^2/\tau^2} + e^{-(t-\delta t)^2/\tau^2}] + A^2 [e^{-(t+2\delta t)^2/\tau^2} + e^{-(t-2\delta t)^2/\tau^2}]$ ; 
$\tau$ = 0.43 ps, $\delta t$ = 0.87 ps, $A$ = 0.52.}
\end{figure*}

Autocorrelation traces of the transmitted seed pulse (normalized to their maximum) are presented in Fig.\ref{Autoco}. They correspond to the integration along the $x$ axis (the beam vertical extension) of the autocorrelation images (top-right image in Fig.\ref{setup}). As a reference, Fig.\ref{Autoco}a presents the case of the output seed after propagation in vacuum (black curve) or in the preformed plasma (blue curve), both without the pump. Fig.\ref{Autoco}b and c present autocorrelation traces of seed pulses amplified at the Joule level ($E_{inc}$ = 40 mJ). One can see that the energy of the pump is transferred within the seed pulse envelope (sub-ps). However, on some shots the autocorrelation trace presented periodic temporal “satellites” (Fig.\ref{Autoco}c) indicating the presence of a main pulse followed by others of lower intensities which could be a signature of a main pulse followed by $\pi$-pulses, as expected from amplification in the self-similar regime \cite{Amiranoff}.
Let us note that in this regime the energy transfer is often associated with pulse temporal compression. However, two dimensional simulations \cite{Trines, Alves, Zhang} show that this compression and the position of the pulse maximum are not uniform transversely, leading to an amplified pulse with a “horseshoe” spatio-temporal intensity profile in the focal region (see 3D-simulations presented in the following). In addition, the traces presented in Fig.\ref{Autoco} are the result of a spatial integration along the transverse (vertical) dimension of the collimated output beam, collected as shown in Fig. \ref{setup}. It was thus impossible to resolve the on-axis pulse duration. Our autocorrelation results are to be intended as an upper-limit of the output seed duration, and the associated output seed peak intensity is likely underestimated.  This is why, for the sake of accuracy, we chose to present the focal spots in Fig. 2 in terms of energy density.

\subsection{D. Raman scattering of the amplified seed.}
\begin{figure}
\resizebox{85mm}{!}{\includegraphics{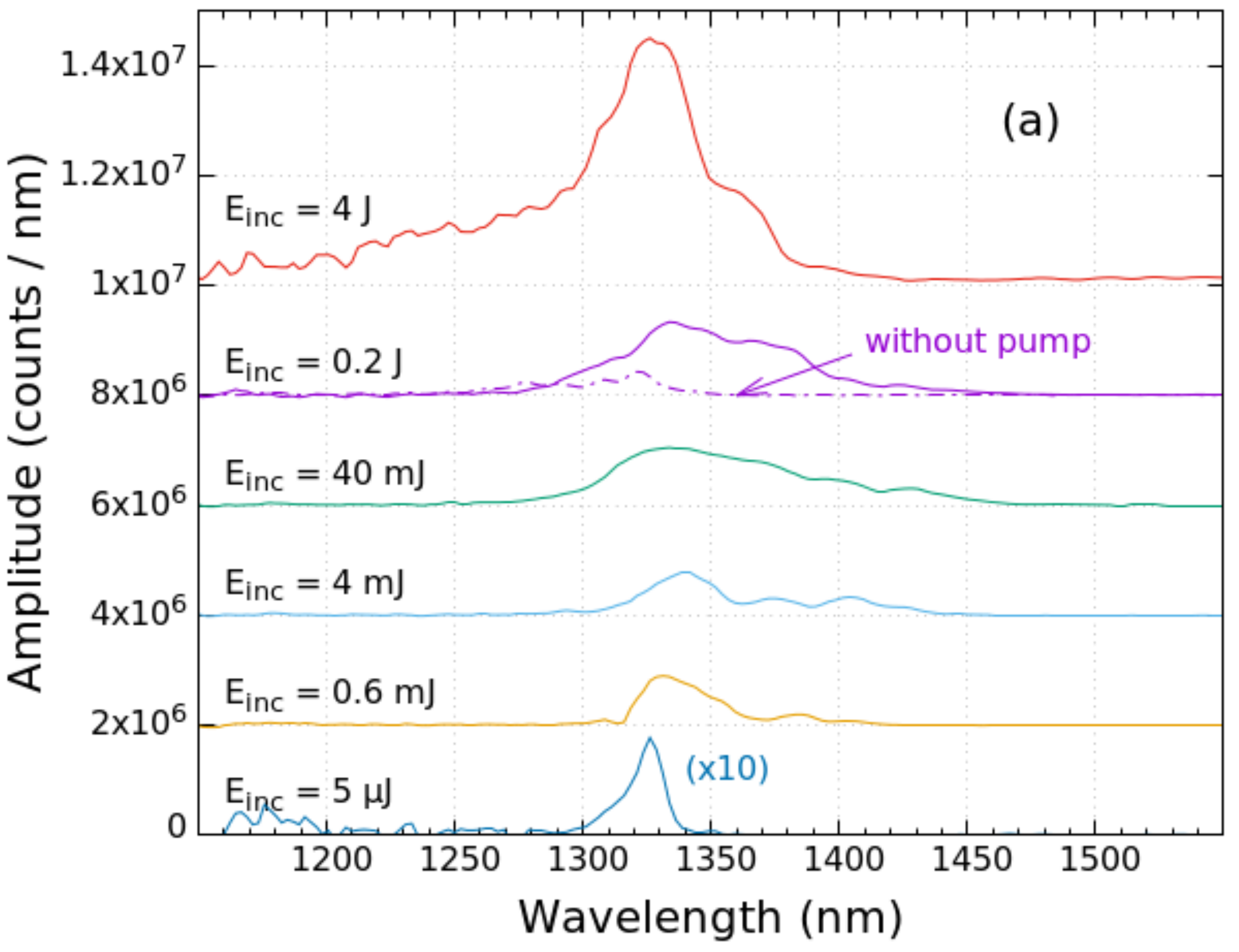}}
\resizebox{85mm}{!}{\includegraphics{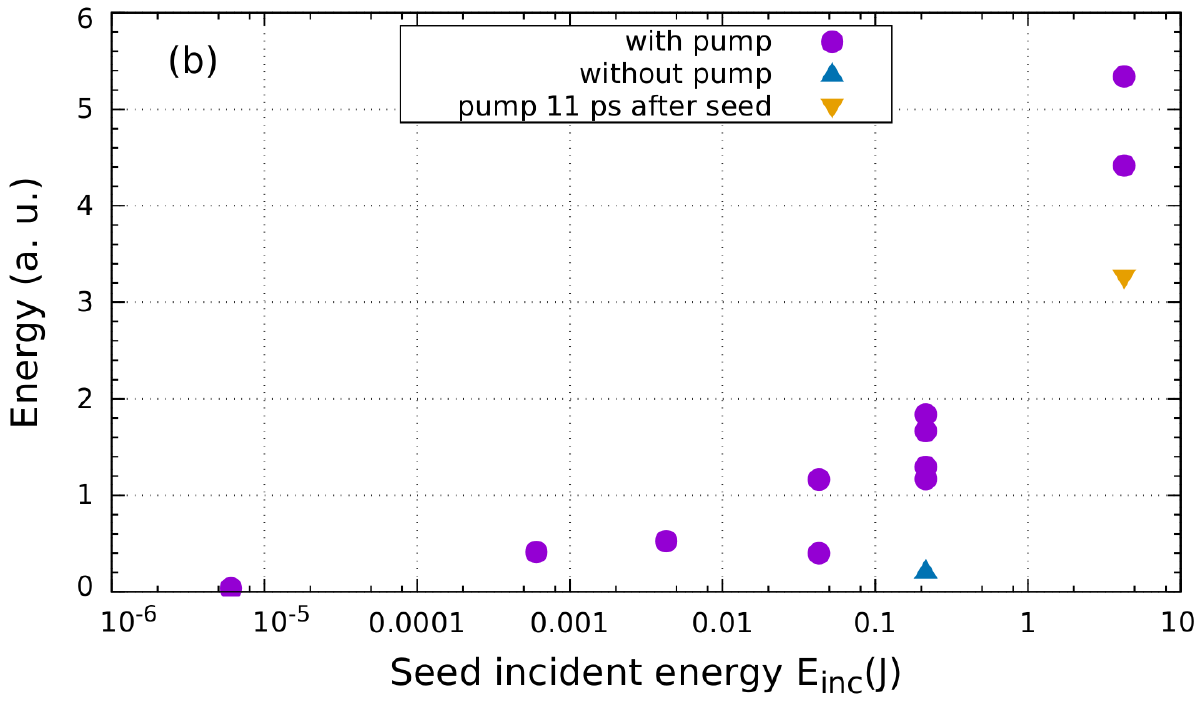}}
\caption{\label{EBRS_vs_Eseed}Seed backward Raman. Spectrum (a) and energy (b) of the seed backward (165$^{\circ}$) Raman at different incident seed energies $E_{inc}$. In fig. (a), for a better visibility, an offset of $2 \times 10^6$ is added to each spectrum, and the spectrum at $E_{inc}$ = 5 $\mu$J is multiplied by 10.}
\end{figure}

Another signature of the increased seed intensity is found in the enhancement of its Raman backscattered signal. Figure \ref{EBRS_vs_Eseed} presents, for different incident seed energies,  the spectra (a) and energy (b) of the seed BRS emitted at 165$^{\circ}$. At low incident seed energy the signal is very weak, indicating that the signal is not coming from forward Raman of the pump. Without the pump, the seed BRS is much weaker (dashed purple curve in (a) and blue triangle in (b) for $E_s^{inc} = 0.2$ J), indicating that most of BRS comes from the intensity increase of the seed. In Fig.\ref{EBRS_vs_Eseed}a one can see that the Raman spectra are centered around 1330 nm and that they never extend closer than 1150 nm toward the pump and seed wavelength side (1058 nm with a FWHM = 5-6 nm), even when $I_{seed}\sim I{pump}$. This allows to exclude a contribution of Raman backscattering to the amplification of the seed.

\subsection{E. Amplification as a function of pump-seed relative delay.}
\begin{figure}
\resizebox{85mm}{!}{\includegraphics{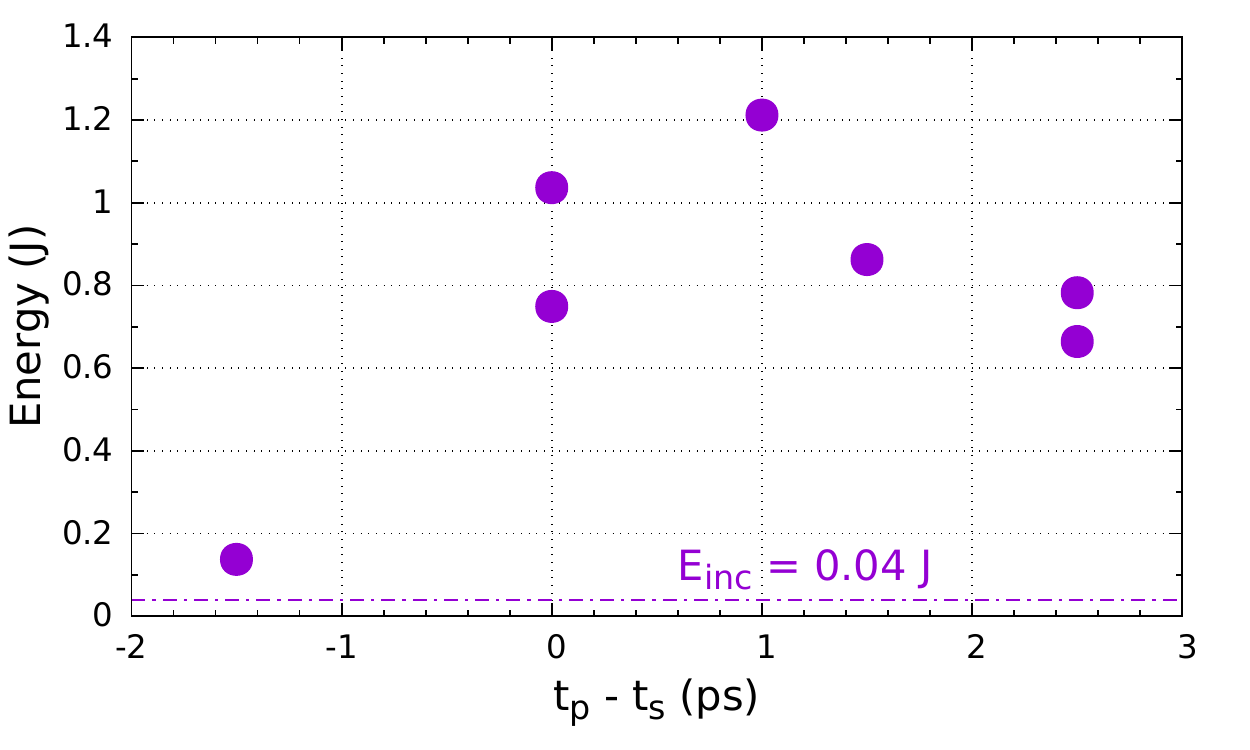}}
\caption{\label{Eout_vs_delay}Seed energy versus pump-seed relative delay. Transmitted energy of the seed as a function of its relative delay with the pump. For $t_p-t_s > 0$ the seed arrives at the plasma center before the pump.}
\end{figure}

The maximum energy transfer not only depends on the incident seed intensity, but also on the pump-seed relative delay $t_p - t_s$, as can be observed in Fig.\ref{Eout_vs_delay}. The maximum energy transfer ($\Delta E\sim$ 1.2 J) is obtained when the seed reaches the center of the plasma slightly before the pump ($t_p - t_s \sim$ +1 ps). Several scans in pump-seed delay at different pump durations and chirp signs (not the purpose of this paper) show this same behaviour.  We explain this by the combination of two effects: i) the amount of spontaneous SRS and SBS undergone by the pump before its interaction with the seed is reduced, preserving the energy available for transfer \cite{Chiaramello2016a}; ii) when the seed arrives earlier it starts to couple to the pump near the maximum of the Gaussian density profile. The low intensity front part of the pump is thus compensated by the high density of the plasma ($\gamma _{sc} \propto (I n_e)^{1/3}$), allowing the self-similar regime to start faster and the coupling to be more efficient \cite{Amiranoff}.

\section{IV. DISCUSSIONS}
In the following we will discuss, with the support of simulations, the different experimental results presented above.\\
\subsection{A. Seed amplification in a three-dimensional geometry.} 
\begin{figure*}
\includegraphics[width=\textwidth]{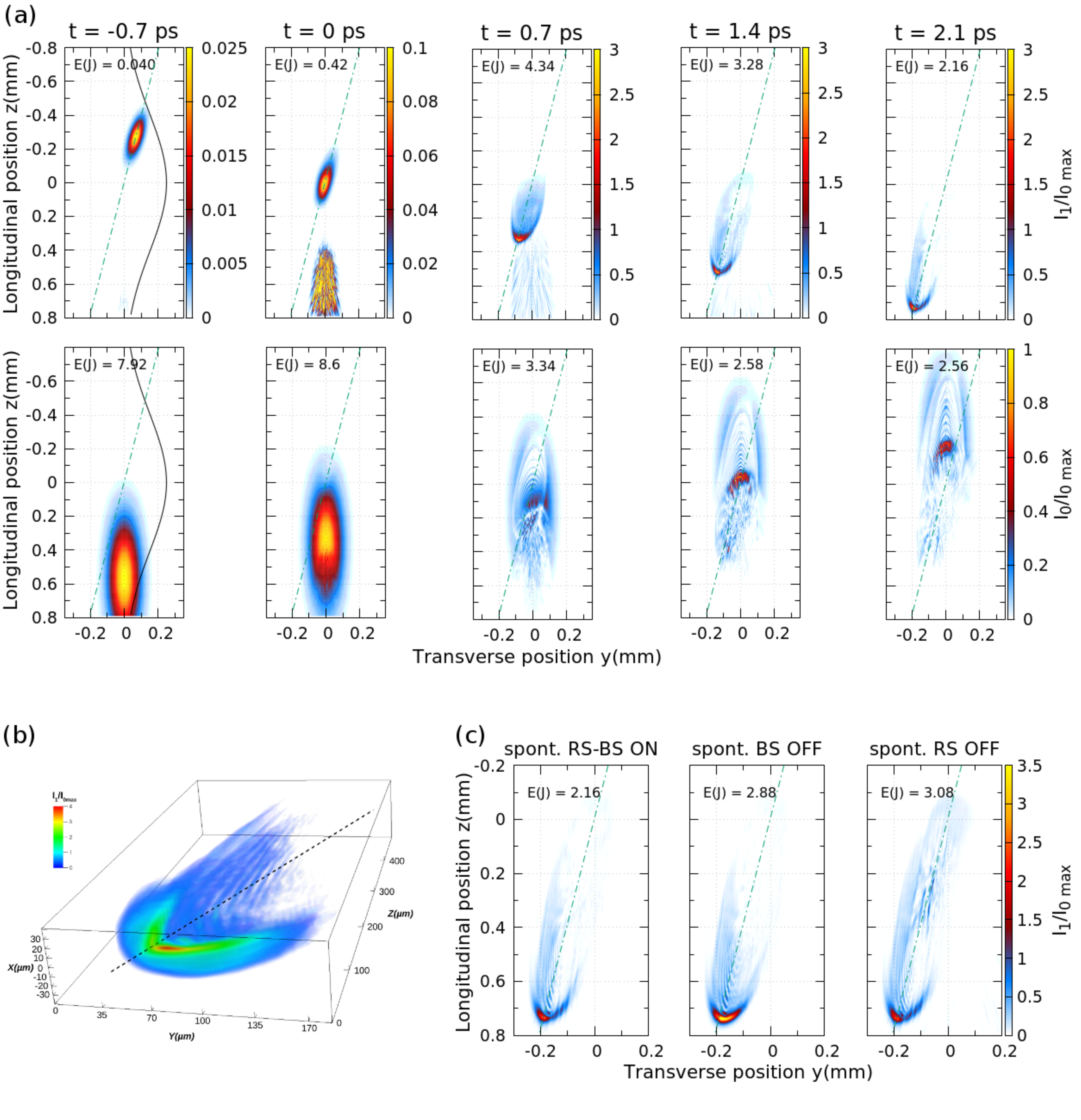}
\caption{\label{Simul_3D}Three-dimensional pump-seed interaction, pF3D simulations: (a): Intensity profiles in the interaction plane ($y$-$z$) of the seed (\textbf{(a)} upper images) and pump ((a) bottom images) pulses at different times (increasing from left to right). The green dashed line indicates the seed propagation axis (165$^{\circ}$ from pump). The black line on the left images indicates the plasma density profile along the $z$-axis. Pump arrives at the plasma center 1 ps after seed ($t_p-t_s$ = +1 ps). Intensities are normalized to the incident pump maximum intensity ($I_{0 max} = 4 \times 10^{16}$ W/cm$^2$). The laser-plasma parameters are the experimental ones, see section Experimental setup. (b): Three-dimensional intensity profile of the amplified seed that reveals the 3D asymmetry (same laser-plasma parameters as in (a), at time $t=$ 1.7 ps (exit of the plasma)). The internal spatio-temporal structure of the pulse is vizualized through the $y$-$z$ and $x$-$z$ plane cuts. (c): Intensity profile in the $y$-$z$ plane of the seed pulse at the plasma exit (t = 2.1 ps), with ((c)-left) or without spontaneous Brillouin ((c)-center) and Raman ((c)-right) instabilities. The central figure ((c)-center) is without ion acoustic fluctuations, SBS always on. The other parameters are the same as in (a).}
\end{figure*}

In order to compare the spatial profiles of the amplified seed beam and the total transferred energy, and in order to study the impact of competing instabilities on the amplification process, we ran simulations using the three-dimensional wave propagation code pF3D \cite{Berger1993, Berger1998}. This code treats each light wave in the paraxial approximation. There are separate wave propagation equations for the pump laser light and the Brillouin and Raman amplified light. The Brillouin and pump light are temporally enveloped at the same frequency. The Raman light frequency is chosen by the user to match the conditions of the application. In addition, the light waves are spatially enveloped along the direction of propagation. The Langmuir and ion acoustic waves (IAW) are also spatially enveloped. However, the IAW is not enveloped in time which allows proper treatment of strongly-coupled SBS. At the maximum plasma density of 0.05 $n_c$ and the maximum pump intensity of $\sim 4\times 10^{16}$ W/cm$^2$ of this experiment, the sc-SBS growth rate is $\gamma_{sc}/\omega_0 \sim 4\times 10^{-3}$ ($\gamma_{sc}^{-1} \sim$ 90 fs), justifying the envelope approximation. Let us note that at this relatively high plasma density, and since both pulses have the same frequency, Raman cannot be considered as a coupling mechanism for amplification \cite{Riconda}. However, spontaneous Raman (and Brillouin) scattering of the pump could reduce its energy before it encounters the seed. The beams spontaneous SRS and SBS are included in the simulations, as well as collisional damping and ponderomotive filamentation. Moreover the effect of the geometry for the beams crossing at an angle (i.e. the effective amplification length in the overlapping region) and the transverse size of the laser beams are fully described. To account for the strong variation of the wave amplitude in the self-similar stage and the 15$^{\circ}$ angle of the seed to this axis, the spatial resolutions are $\Delta x = \Delta z = \lambda_0$ and $\Delta y = 2\lambda_0 /3$. Tests were made to make sure resolution was adequate. In addition, the axis of propagation was tilted by 7.5$^{\circ}$ so both pump and seed propagated at $\pm$ 7.5$^{\circ}$ to the axis with the same results. The light waves interact with the background plasma through ponderomotive forces which can intensify the light and cause further development of small spatial scales. Because of the short time duration, low plasma density, and low charge state ($Z$ = 1), filamentation was not found to be important, it will not be further discussed. To take into account nonlinear effects on the ion perturbation, such as wavebreaking, and in accordance with typical PIC simulation results (see following), its amplitude was limited to 50\%. This does not affect the amplification of the main part of the seed pulse, for which the density perturbation stays below this limit, but prevents a non-physical scattering of the pump pulse on these perturbations long after the seed pulse has interacted with the pump \cite{Andreev2006}.\\
The input laser and plasma parameters are those given in the Experimental setup paragraph. The spatial and temporal profiles are Gaussian. The hydrogen plasma density has a peak value of $n_e/n_c$ = 0.05, with a FWHM = 575 (750) $\mu$m along the $z(y)$-axis respectively and is constant along the $x$-axis. The ion temperature is $T_i$ = 80 eV. The electron temperature is $T_e$ = 100 eV for pF3D. Pump and seed wavelength is 1057 nm, their durations are respectively 1.7 ps and 0.55 ps (FWHM). Their focal spot sizes (FWHM) $\sigma_x \times \sigma_y \simeq 85\times 135\, \mu{\rm m}^2$ and $\sigma_x \times \sigma_y \simeq 100\times 36\, \mu{\rm m}^2$. The pump peak intensity was $4 \times 10^{16}$, corresponding to an energy of 9 J. The seed peak intensity was $10^{15}$ W/cm$^2$, corresponding to $E_{seed}$ = 40 mJ in Fig. \ref{Eout_vs_Eseed}.\\
Figure \ref{Simul_3D}(a) shows, at five different interaction stages, the spatial profiles of the pump and seed electric fields in the interaction plane (y-z, cf. Fig. \ref{setup}). At $t$ = -0.7 ps the seed pulse is in the plasma, it has not yet encountered the pump pulse, and its initial energy (40 mJ) and intensity are preserved (no significant losses). At $t$ = 0 the peak of the seed is in the middle of the plasma ($z$ = 0), where the pump peak arrives 1 ps later ($t_p-t_s =$ +1 ps). The pump front however already starts to transfer its energy and the peak intensity of the seed has increased by a factor 4 ($I_1/I_{0 max} \sim 0.1$). At $0.4<z<0.8$ mm (pump plasma entrance) one can see scattered light. This is spontaneous backward Brillouin of the pump, that starts to deplete its trailing edge. At $t$ = 0.7 ps the seed peak intensity is 120 times larger, reaching 3 times the incident pump one ($I_1/I_{0 max} \sim 3$). The pump profile shows a strong depletion with two patterns: one at its front dug by the seed (stimulated Brillouin), and an inhomogenous one induced by spontaneous backward Brillouin at its back. Due to the 165$^{\circ}$ interaction geometry, the seed front tends to bend and expand toward the pump front. The pump-seed overlap ends at $t$ = 1.4 ps. When the seed leaves the plasma ($t$ = 2.1 ps) it energy has reached 2.16 Joules, close to the measured one (Fig. \ref{Eout_vs_Eseed}). After the interaction the pump has undergone a strong depletion (70 \%), close to what we measured (Fig. \ref{Eout_vs_Eseed}). Let us note that most of the energy transfer occurs within $\sim$ 400 $\mu$m.

Figure \ref{Simul_3D}(b) shows a 3D view of the amplified seed, revealing the 3D asymmetry. It corresponds to the same simulation as the 2D slices presented in Fig. \ref{Simul_3D}(a), at time $t=$ 1.7 ps (exit of the plasma). The internal spatio-temporal structure of the pulse is vizualized through the $y$-$z$ and $x$-$z$ plane cuts.\\
It is important to point out that although the level of spontaneous scattering is significant, it is not enough by itself to justify the low transmission of the pump alone in our experiment, where other factors (discussed in section III-A) need to be taken into account. Thus, decreasing spontaneous scattering and increasing the quality of the created plasma are both important points to consider in order to increase the pump transmission and the energy available. The role of spontaneous instabilities and how to reduce them is discussed in the following.

\subsection{B. Influence of spontaneous pump backscattering.} Spontaneous Brillouin and Raman scattering have a significant influence on the results. Because of its faster growth rate Raman is the most important. Due to an initial temperature $T_e \approx 100$ eV , $k_0\lambda_{De}$ = 0.063 and, with $k_{lw} \sim 1.6 k_0$ , $k_{lw}\lambda_{De}$ = 0.1 so Landau damping is unimportant at the peak density of 0.05 $n_c$ ($k_0$ and $k_{lw}$ are the wavenumbers of the laser and the Langmuir wave respectively, and $\lambda_{De}$ is the Debye length). Even at a lower density of 0.01 $n_c$ , $k_{lw}\lambda_{De}$ = 0.22. pF3D simulations show that the plasma is heated to $T_e$ as high as 250 eV when Raman is included. Nonetheless, Raman reaches saturation too fast for heating to affect the results. To study the influence of spontaneous Raman and Brillouin on the amplification process, we have run 3D simulations with the SRS or the ion density noise turned off (stimulated Brillouin always on). This comparison is interesting even if a realistic treatment of the plasma fluctuations in envelope codes is not trivial, while in PIC codes the noise level is exagerated and spontanous backscattering overestimated on the short time scales used here. Fig.\ref{Simul_3D}(c) shows the spatio-temporal profile (intensity normalized to the pump maximum one) of the seed at the plasma exit obtained with pF3D for three cases: with spontaneous Raman and Brillouin (\ref{Simul_3D}(c)-left, same as fig. \ref{Simul_3D}(a), without Brillouin (\ref{Simul_3D}(c)-center), without Raman (\ref{Simul_3D}(c)-right).
The differences are explained as follows: At the plasma entrance, the ion density grating generated by spontaneous Brillouin scatters the pump light in its backward direction. At the plasma center, the angle between the two beams leads to a density grating that scatters the pump in the seed direction. Without spontaneous Raman, the ion grating is stronger and more energy is left at the back of the pump to be scattered in the seed direction, which forms the long tail observed in fig. \ref{Simul_3D}(c)-right.
Without spontaneous Brillouin, the ion grating has a more homogenous spatial profile, which leads to a more homogenous intensity profile of the amplified seed. Also, one can observe that the amplification process in crossed geometry tends to bend and extend the intensity distribution toward the pump axis ($y$ = 0), as observed experimentally (fig. \ref{Foc_seed}). This is induced by the combination of several factors.
We have shown previously \cite{Amiranoff, Chiaramello2016a} that the evolution of the phase that governs the three waves interaction makes scSBA more efficient if the seed propagates in a negative density gradient (from high to low density), so in our case in the $z>0$ side of the Gaussian density profile. We also observed (Fig. \ref{Eout_vs_delay}) that, by reducing the spontaneous pump scattering and thus preserving the pump energy, the maximum energy transfer occurs for pump-seed delay $t_p-t_s > 0$ (+1 ps in Fig. \ref{Simul_3D}(a)), that is for a seed that has already passed the plasma density peak before it overlaps the pump peak (again the $z>0$ region).
Finally, an important result of this paper (see figure \ref{Eout_vs_Eseed}) is that significant energy exchange can still occur at low seed incident intensity, so even the wings of the seed can be amplified. Combining these effects with the not-counter-propagating geometry ($165^{\circ}$) leads to an asymmetric amplification process in favor of the $z>0$ region, with an energy transfer that tends to be larger toward $y=0$, where the pump intensity is maximum.

\subsection{C. Influence of the incident seed intensity.} 
We performed a large number of simulations in order to study the dependence of the energy transfer on the seed incident intensity. The pF3D envelope simulations were completed by PIC simulations with the code SMILEI \cite{Smilei} in a one-dimensional geometry. In the latter, even if the  165$^{\circ}$ geometry of the interaction is not simulated, ion and electron nonlinearities (such as harmonics, trapping or wavebreaking) are included as opposed to the envelope code. Moreover spontaneous Brillouin and Raman scattering are included in a self-consistent way. The pF3D and Smilei simulations have been run with the same laser-plasma input parameters, the experiment ones (see section Experimental setup). In the Smilei simulations the experimental pump frequency chirp ($\alpha = -7.7\times 10^{-7}$) was included, even if the results of the simulations indicate that it does not significantly modify the energy transfer in the present experiment. To take into account the heating by inverse bremsstrahlung absorption of the pump simulated in pF3D, $T_e$ is set at 300 eV in the 1D PIC simulations with the pump, and 100 eV without. The PIC simulations were run with 300 particles/cell, giving at $T_e$ = 300(100) eV a the spatial resolution better than 2.5(5) Debye length. The time step was $dt$ = 0.175 fs (0.95 $dx/c$). The initial positions of the pump and seed Gaussian pulses were adjusted depending on their relative delay $t_p-t_s$, the intensity of the rising edge at the entrance of the simulation box never exceeding $1/e^2$ of their respective peak intensity. The boundary conditions were reflective for the particles and open (``silver-muller'') for the electro-magnetic waves.

\begin{figure*}
\includegraphics[width=\textwidth]{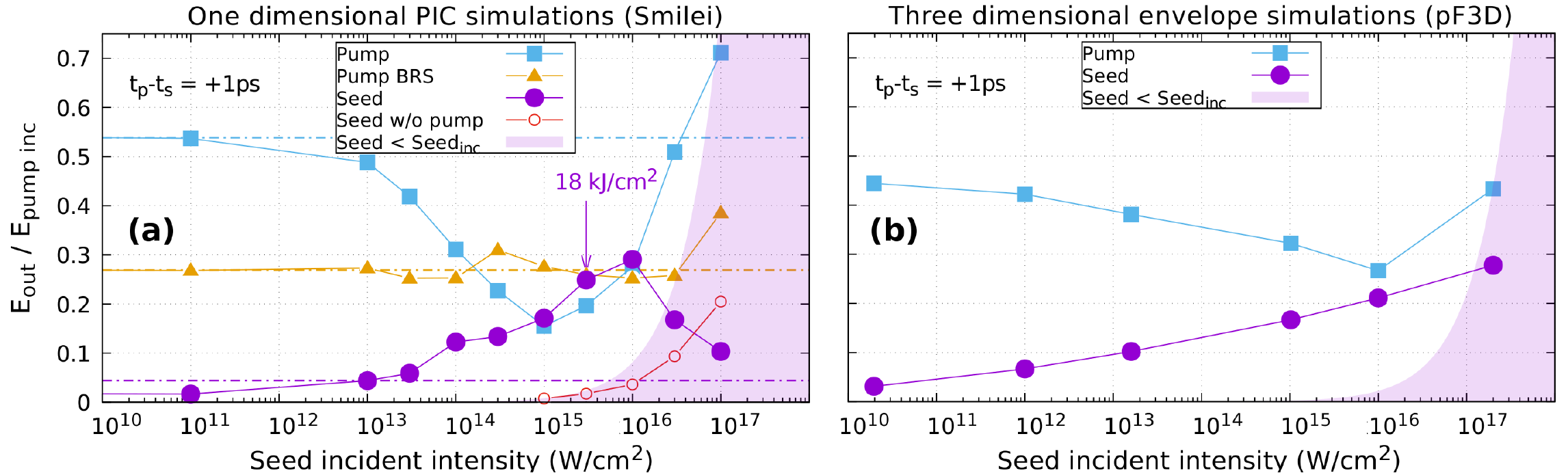}
\caption{\label{Simul_vs_Iseed}Simulations, output energies versus seed incident intensity. (a): 1D-PIC (Smilei) and (b): 3D-envelope (pF3D) simulations, with the same laser-parameters as in Fig. \ref{Simul_3D}(a). The energies at the plasma exit of the pump, seed, as well as the pump backward Raman scattered light, are normalized to the incident pump one. The pump-seed delay is $t_p-t_s$ = +1 ps. The filled areas correspond to a seed output energy lower than the incident one (energy transfer from seed to pump). The dashed lines in (a) show the energies without seed of spontaneous Brillouin, Raman, and transmitted pump (from bottom to top respectively).}
\end{figure*}

Fig.\ref{Simul_vs_Iseed} shows the pump and seed energies at the plasma exit as a function of the seed incident intensity, for 1D-PIC (left) and 3D-envelope (pF3D, right) simulations, at $t_p-t_s =$ +1ps. In both cases, one can observe the same behaviour as the experimental results presented in Fig. \ref{Eout_vs_Eseed}: an increase of the transferred energy from pump to seed when the incident intensity is increased, up to an optimum. When the incident intensity is larger than the pump one, the initial conditions for optimum amplification are not met and the efficiency drops. In particular, the seed duration is too long for this configuration \cite{Chiaramello2016a}. 

The dependency of the energy transfer on the incident seed intensity can be explained as follows \cite{Amiranoff}. During the coupling between the pump beam and the seed beam, the system successively explores different stages. In the initial stage, the phase of the seed adapts to the pump but the energy transfer between pump and seed is negligible. 
In the second stage, both the seed and the ion wave grow exponentially and the energy exchange starts to be important when $I_{seed}\sim I_{pump}$. Pump depletion then becomes significant and in this so-called self-similar regime, most of the remaining pump energy is transferred to the seed. The relevant parameter is thus the time and location at which pump depletion sets in. 
Indeed, from this time, the seed stores all the remaining part of the pump pulse: {\it{the earlier the pump depletion, the larger the energy gain}}. The larger its incident intensity, the faster the seed enters the exponential stage and grows to an intensity comparable to the pump, thus maximizing the energy gain. However, when $I_{seed}$ is initially equal to $I_{pump}$, the energy transfer from pump to seed only takes place in a very small zone (wings of the seed), and an efficient gain requires a longer plasma and a longer pump duration. Notice that amplification of seeds more intense than the pump can be obtained from the beginning of the crossing in a different regime \cite{Riconda, Weber}.

In 3D,  the interaction of the wings of the pump and seed beams occurs at lower intensity than on axis, so that, on average, the maximum seed intensity at which the energy transfer is reversed (seed to pump) is larger than in 1D. From the 1D PIC simulations one can see that without the seed, or at low seed intensity, the spontaneous backward Raman and Brillouin scattering of the pump represent respectively 30 and 5 \% of the incident pump energy.
However, when the incident seed intensity is increased, spontaneous Raman remains unchanged while the energy transfer due to SBS becomes significant, up to 30 \% of the initial pump energy. Complementary simulations have been performed increasing the ion mass by 100. They confirm that the ions are the vector of the energy transfer from pump to seed: at an incident seed intensity of $10^{15}$ W/cm$^2$ and $t_p-t_s$ = 0 the output seed has an energy 5.5 times lower and is twice longer. Accordingly the transmitted pump energy is 1.8 times larger, while the Raman signal is almost unchanged.
 
In conclusion, both 1D and 3D approaches show the same behaviour of the amplification process with the incident seed intensity, and are in good agreement with the experimental results of Fig. \ref{Eout_vs_Eseed}. As an example, the energy per unit area obtained with the 1D PIC code (18 kJ/cm$^2$ for $I_{seed}= 3\times 10^{15}$ W/cm$^2$) is in excellent agreement with the measured one. In addition, the relatively slow increase of the energy transfer with the incident seed intensity explains the increase of the spatial width of the amplified seed: the relative gain $\Delta E/E_{inc}$ of the low-intensity wings is larger than that of the high-intensity central part of the beam.\\

\subsection{D. Backward Raman scattering of the amplified seed.} 
\begin{figure}
\resizebox{85mm}{!}{\includegraphics{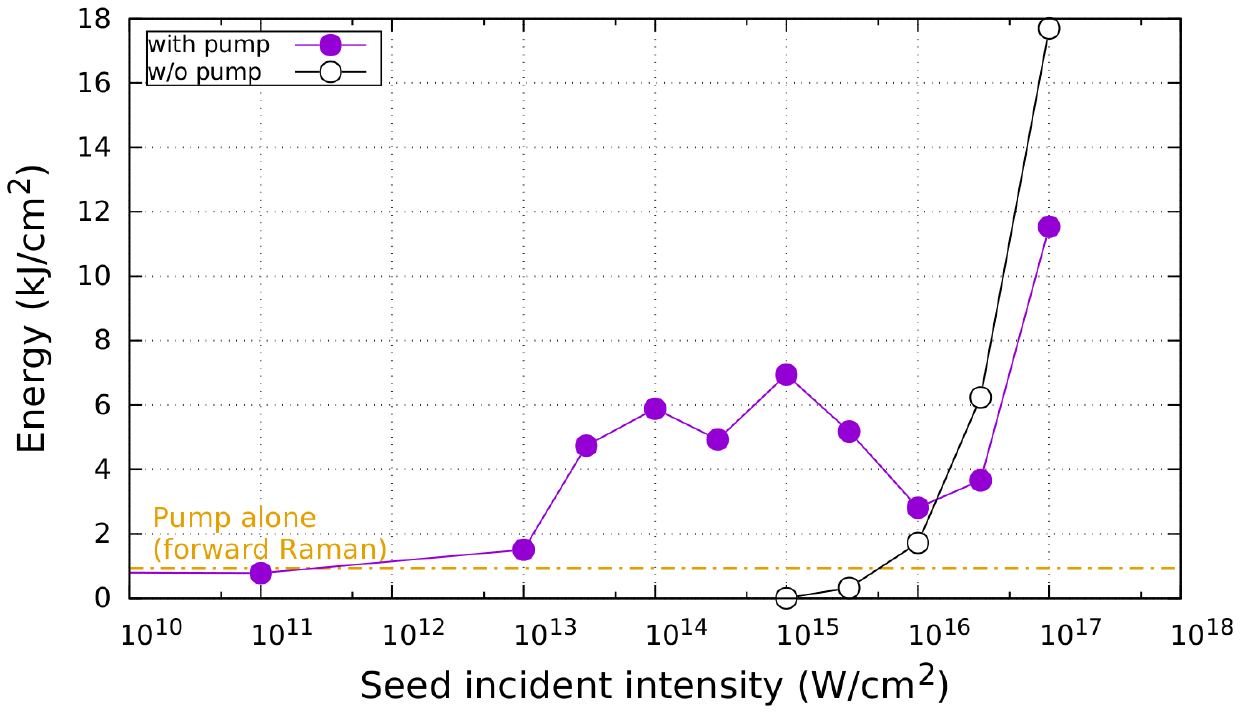}}
\caption{\label{Smilei_Seed_BRS_vs_Iseed}1D PIC simulations, backward Raman energy from the seed. The dashed orange line is the level of Raman without seed, generated by forward Raman of the pump. The pump-seed delay is $t_p-t_s$ = +1 ps.}
\end{figure}

In order to compare with the experimental results of Fig.\ref{EBRS_vs_Eseed}, the Raman energy propagating backward from the seed (forward from the pump) is presented in Fig.\ref{Smilei_Seed_BRS_vs_Iseed}, without or with the pump pulse, for the same set of simulations reported in figure \ref{Simul_vs_Iseed}-left. Almost no signal is generated when the pump is alone or the seed very weak. Without the pump (black curve), Raman becomes non-negligible for incident seed intensities near or above $10^{16}$ W/cm$^2$. When the pump is on and the incident seed intensity is between $10^{14}$ to $10^{16}$ W/cm$^2$, the Raman signal increases, in correlation with the seed amplification observed in Fig. \ref{Simul_vs_Iseed}. At an incident intensity of $10^{15}$ W/cm$^2$ the signal is 1500 times larger, equivalent to the signal observed with the seed alone at an intensity 60 times larger (black curve). On the contrary, for an incident seed intensity above $10^{16}$ W/cm$^2$, the seed will transfer energy to the pump thus decreasing the spontaneous seed backscattering.

\section{CONCLUSIONS AND PERSPECTIVES}
In this paper laser-plasma amplification of sub-picosecond pulses above the Joule level was demonstrated, a major milestone for this scheme to become a solution for the next-generation of ultra-high intensity lasers. We demonstrated the importance of using a seed pulse of initial intensity high-enough to enter the self-similar regime as promptly as possible, and get a highly efficient (20 \%) energy transfer in a sub-picosecond scale. In the regime of efficient amplification, evidence of energy losses of the seed by spontaneous backward Raman was found. Simulations with a three-dimensional envelope code, supplemented with a one-dimensional particle-in-cell code, were compared with the experimental results. They show very good agreement with the amount of transferred energy, and with the global behavior observed in the experiment. Spontaneous Raman, as well as spontaneous and seeded Brillouin scattering reach nonlinear levels. The pump transverse intensity profile, together with the crossed beam geometry, cause the seed intensity distribution to bend and extend toward the pump axis. This could be avoided (limited) by using a pump beam with a larger focal spot or with a tilted intensity front \cite{Akturk} adjusted to the crossing angle.
The output seed energy, intensity and quality were limited by the homogeneity and extent of the spatio-temporal interaction volume, itself limited by the energy and quality of the laser beams. Without reducing the scSBS gain, the beam losses from spontaneous Raman scattering and collisional damping might be limited with the use of a plasma with a density ramp and a much higher temperature. An initially shorter seed pulse will also reduce these losses, and would be better adjusted to the self-similar regime, increasing the pulse compression \cite{{Chiaramello2016a},{Trines},{Trines2018}}. Much larger energy transfers might be reached using longer pump and plasma, larger focal spots of homogeneous and hyper-Gaussian profiles. Exploiting more energetic lasers will allow these improvements and open the way to the numerically predicted \cite{Riconda}\cite{Weber} next step: an output seed intensity much larger than the pump one.

\section{ACKNOWLEDGEMENTS}
We greatfully acknowledge the ELFIE technical staff for the invaluable help during the experiments, Mickael Grech, Frédéric Perez and Tommaso Vinci for their help on the PIC code Smilei, Pascal Loiseau for fruitful discussions and 3D fluid simulations of plasma creation and heating by the ionization beam. This work has been done within the LABEX Plas@par project, and received financial state aid managed by the Agence Nationale de la Recherche, as part of the programme "Investissements d'avenir" under the reference ANR-11-IDEX-0004-02. S. Weber was supported by the projects High Field Initiative (CZ.02.1.01/0.0/0.0/15\_003/0000449) and ADONIS (Advanced research using high intensity laser produced photons and particles, CZ.02.1.01/0.0/0.0/16\_019/0000789), from European Regional Development Fund (HIFI). M. Blecher. was supported by a grant from GSI. The research leading to these results has received funding from LASERLAB-EUROPE, grant agreement n\textsuperscript{o} 284464, EC's Seventh Framework Programme. L. Lancia was supported by CRISP, contract n\textsuperscript{o} 283765, EC's Seventh Framework Programme.

\end{document}